# Interview with Warren Wiscombe on scientific programing and his contributions to atmospheric science tool making.

Piotr J. Flatau

**Abstract**. On March 11, 2013 I talked with Warren Wiscombe about his contributions to scientific computer programing, atmospheric science and radiative transfer. Our conversation is divided into three parts related to light scattering, radiative transfer and his general thoughts about scientific programing. There are some reflections on how radiative transfer parameterizations gradually sneaked in to modern Global Circulation Models. Why some software programs such as light scattering code MIEV and DISORT are very successful and why some of them fizzle. We talked about the role of tools in modern science, open source movement, repeatability of scientific results, computer languages, computer programs as objects of arts, and even if programs can be revolutionary.

This report can be cited us:

Flatau, P. J. 2013, Interview with Warren Wiscombe on scientific programming and his contributions to atmospheric science tool making, http://arxiv.org/abs/1304.1582

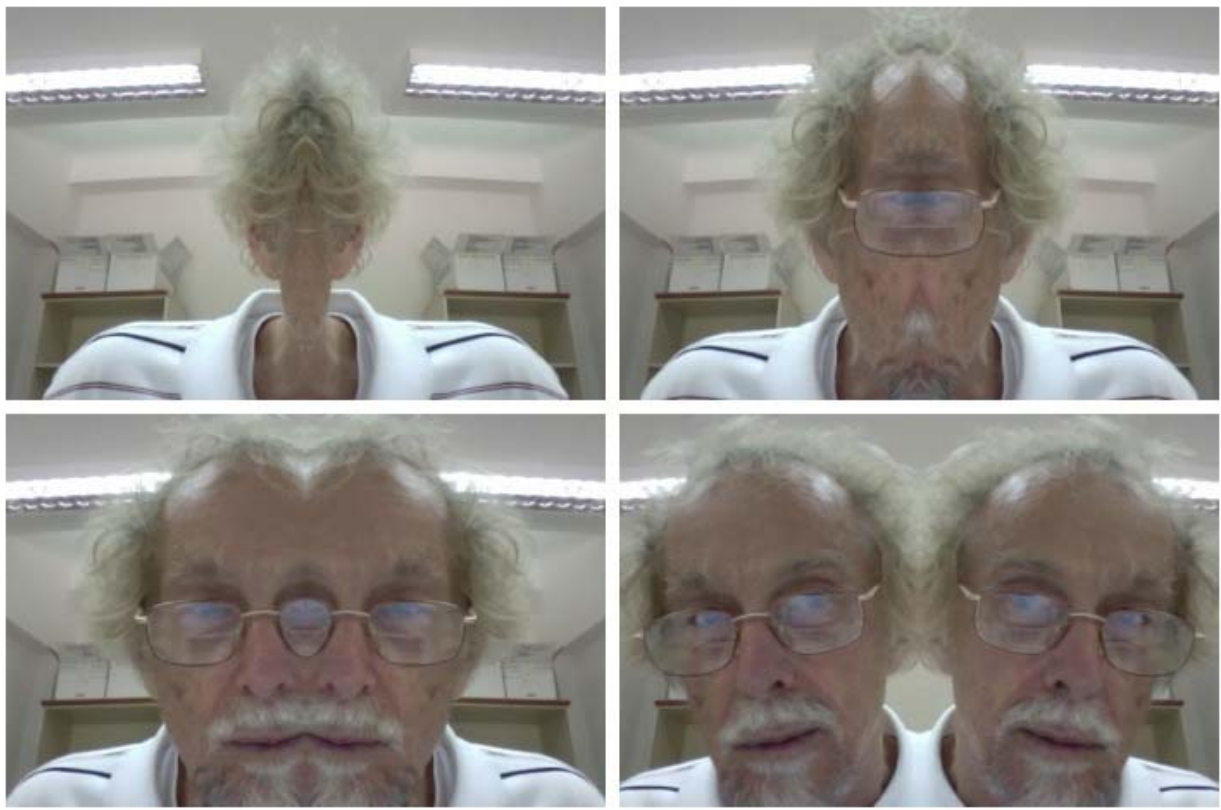

Warren Wiscombe

# Light scattering

[Piotr J. Flatau] **You know Warren I still have a folder of your publications which I started to collect when I was a student at Colorado State University. You are one of a few researchers whose papers I've really read and I was studying your codes. I divided the talk into several sections: light scattering, radiative transfer and general issues. I do not think that we will be able to go over all of them. But let us begin. In January of 1973 you wrote a letter to Colonel John Perry. In this letter you say that it would be beneficial to provide Mie calculations in the form of tables which would be accessible to general community. Before you van de Hulst and others struggled with Mie calculations. In that time it was a difficult problem. I would like you to set the stage and describe the status of light scattering calculations in the 1970s. I would like people to understand how heroic it was at that time.**

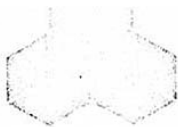
# SYSTEMS, SCIENCE AND SOFTWARE

January 19, 1973
AIR MAIL

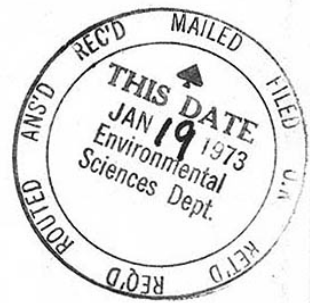

Colonel John Perry
DARPA/IPT
1400 Wilson Boulevard
Arlington, VA 22209

Dear Colonel Perry:

There are a large number of research groups in this country (many of whom participate in the DARPA network) who perform Mie scattering calculations. There is great interest in problems of radiative transfer through aerosols and clouds, both in the technical community and in the community at large, because of their potential climatic impact (SCEP report, SMIC report, etc.). Therefore, the quantity of Mie calculations performed in the future will most likely escalate.

This would be of little concern if such calculations were inexpensive, but, in fact, Mie computations are notorious for the quantity of computer time which they gobble up. There is, therefore, a tremendous wastage of computing funds (almost entirely at government expense) when, as is the case, identical computations are performed at many different facilities.

This waste could, to a large extent, by alleviated by providing central tables of Mie quantities available throughout the DARPA network. We have made estimates of the amount of data such tables would need to contain, and it runs to many billions of words, far beyond the storage capacity of current-generation computers. Probably the only feasible place for such tables to reside would be the laser store of ILLIAC IV. I believe, after numerous conversations with people in the field, that such tables would be heavily and productively used, for Mie computations are now so standardized that there is no glory for a researcher in

> Colonel John Perry        -2-        January 19, 1973
>
> writing his own code — it is merely a chore, and a costly one as far as the government is concerned.
>
> I understand that ILLIAC IV is proceeding considerably behind schedule. However, I do not believe that preliminary studies on the creation of Mie tables are premature. A consensus from authorities in the field on the exact structure of such tables will be needed, and the computer code to generate the tables can be written in advance of the actual availability of ILLIAC IV. Naturally, since we have made numerous Mie computations ourselves and are familiar with the radiation community, we believe that we are eminently qualified to undertake the task.
>
> If this idea interests you, we would be happy to discuss it further by telephone.
>
> Cheers,
>
> Warren J. Wiscombe 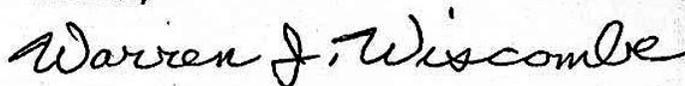
> Environmental Sciences Dept.
>
> WJW:prr
>
> cc: Dr. Lawrence Roberts
>     Director, IPT

[Warren Wiscombe] Right, Perry was the head of ARPA climate dynamics program. It is amazing that you've got this. I am not sure I have it.

**You do.**

Of course the community was very small and everyone used in that time Dave's codes. He was an odd bird at the IBM Research Lab, in the days when IBM would support a lot of research at their lab. We were happy to use Dave's codes at the beginning. They were certainly better than anything available; they were the only codes available. He did a very careful job of looking at where you cut off series that you are summing and make sure that you have accurate results and so forth.  Over the years that I used his codes I kept finding areas where, when you poked too hard, the code would break down.  I kept the list of the things and I have fixed them as time allowed. I developed a log of fixes of the Dave code. His coding was opaque at best, although typical of the times. I tried to develop a more flowing style that was elegant and well documented both inside and outside the code. Dave's code was well documented outside of the code, he published IBM reports and I followed his example in that respect for both Mie codes and DISORT.  But it reached the point when I felt that the Mie code he distributed was simply not adequate for the kind of work problems that we were attacking. It wasn't that it was slow, computers were slow in those days, and you could spend hours doing Mie calculations when you integrated over the drops sizes in the cloud, for example. It was simply that sometimes he summed his series too far and he would add terms that would be wrong. Mie series is kind of weird series. It doesn't actually converge. It converges, and then diverges when you keep going because calculations you use to calculate some of the functions there, some of the Bessel functions, start to diverge. One has to cut it off very precisely. In fact one of the results people love the most of the improvements I made was that I have said that the Mie code should stop at $x+4$ (x to the 1/3 power). That result propagated around the world. Everybody was using it. In fact, it was from some work that Nussenzweig had done.

# SUBROUTINES FOR COMPUTING THE PARAMETERS OF THE ELECTROMAGNETIC RADIATION SCATTERED BY A SPHERE

J. V. Dave

IBM Scientific Center
Palo Alto, California

## ABSTRACT

Two experimental FORTRAN subroutines with which one can compute the so-called "Efficiency Factors" and the Stokes parameters of the electromagnetic radiation scattered by a sphere, are described in great detail. The index of refraction (m) of the material of the sphere is assumed to have the form $n_1 - i\, n_2$ with $n_2 \geq 0$. The formulas used for these subroutines were first derived by G. Mie (1908) and as such this scattering process is referred to as Mie scattering in the scientific literature.

A listing of these subroutines along with sample program outputs are provided in the appendices.

**Yes, I remember. After you there was a book by Bohren and Huffman and they used it.**

It became universal criterion. It was one of many improvements I have made. Some of them were smaller, less noticeable, but they were important. Like, Dave's code did not work as the refractive index approached one, and yet bacteria were in that regime. You did not want to have code which failed in that regime if you were biologist. That is the story of the Mie code. He published his in 1969 and I published mine in 1979 both as a paper and as a report and I tried to use ideas and scientific software of the time and a little bit of software engineering although that came around much more strongly later. I became aware of this vast software engineering literature that wasn't all stupid. Scientists have a tendency to think that it is stupid that they are just twiddling knobs, and it is not really important, but actually I found number of very important ideas in that community and I brought it back. But I think that in the time of the Mie code, say in 1979, I haven't brought a lot of these ideas. I kind of developed my own style. Things were well documented, there were a lot of comments, the code writing was not opaque, and one could pretty much follow it. I was on my way towards the style that I have eventually perfected. The tables, it looks silly now. But you may remember back in late 60's and early 70's everybody was using tables from the reports by Deirmendjian [Dei1962]. Do you remember the Deirmendjian C.1 Cloud?

**Yes I remember. I also remember seeing a huge book in the library, it was newspaper size.**

Yeah. These were Coulson [Coulson1960] tables for Rayleigh scattering including polarization. Those were enormous. But the Deirmendjian book was smaller but it had tables of phase functions. Everybody was using phase function Haze L and Cloud C.1; or they used the Henyey-Greenstein function which came from astrophysics. Nobody paid much attention to Mie calculations simply because they were too damn expensive. If you put a realistic drop distribution with cloud drops going up to 30 microns in size you would be looking at 1/3 to 1 hour computer time to get Mie properties, for 200 wavelengths in shortwave. It was serious computation. Now, you snap your fingers, this is how fast it is done. Another thing about Mie computations were resonances, those spikes that occur. Those were also scary because if you happen to land on one it would really distort your integral over the sizes. Of course, you can hardly avoid landing on resonances because they are all over the place. People would do these integrations over the drop sizes and would then double the number of points for the quadrature and they would get different results and they would quadruple the number of points and get another result and so on and so on and it would never converge. The reason you do not converge is that you land on resonances, maybe not right on the peak but on the shoulder. It was very maddening for me. I remember tearing up sheets of computer paper. We never got it to converge properly, not in any mathematician sense.

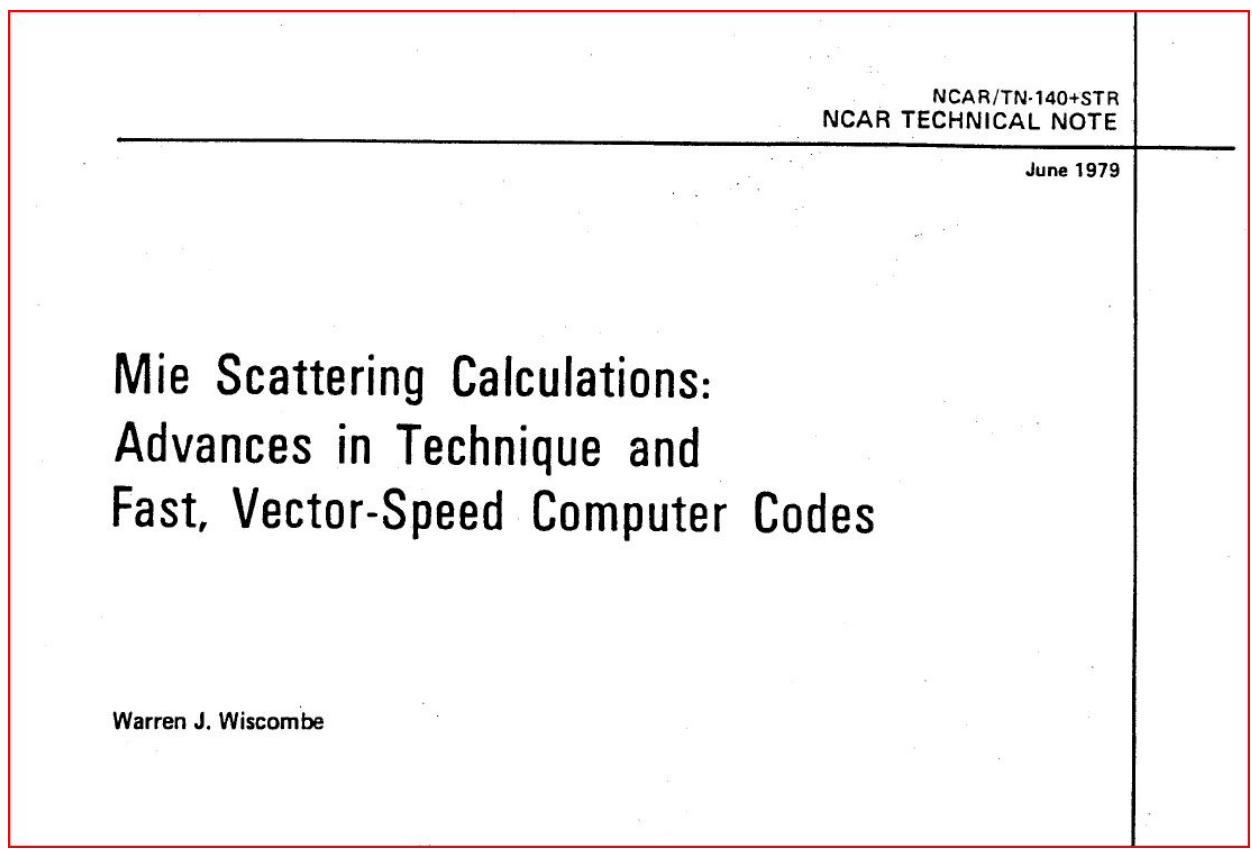

**This was all at NCAR?**

Yes it was all done at NCAR. It was all sideline activity for me. I was really beating on the doubling method and calculations such as delta Eddington at that time.

**What was the NCAR environment as far as supporting coding?**

NCAR was like a dream. It was like you wish that scientific institutions were like that. It is not like that anymore. You had your freedom; you could do what you wanted. They were all these facilities available including this CRAY computer which we all were mesmerized by because we could vectorize our code and speed it up by a factor of seven. We all learned how to write a code so it would vectorize properly.

**Did you have help at NCAR who would help you programming or it was you who were the leader in that time?**

Certainly the people who were supporting the core NCAR software they were aware of these issues. But I think rather dimly. I was not their leader. But I think I was leading in terms of ideas, in the level of detail that you need to go into to document scientific software properly. It is not enough to just list variable names. I would add long paragraphs of text into documentation and say – look here is where you can get into trouble, don't worry about these variables, I tried to get some insight, so people where not just swimming with the sharks, they actually had some guidance or even do something about the code. That was what was missing in those days. The kind of documentation they were doing was cold and sterile; it didn't have any human touch to it. It took big modelers a long time to get into documenting their codes. They simply felt that they were changing the code all the time so there was not a point of documenting anything. I later realized that there was just a total disconnect with software engineering. In software engineering of course, whether it is large code such as those which run a telephone system or small ones - these codes they are changed too, but they still are documented and they have configuration control and they are very careful about changes. The more frequently you change the code the longer it will take to perfect it. In those days modelers did not believe in it or know about it. Everybody was madly typing at key punch machines, changing their computer cards all the time. Models were moving targets, so why document them? It wasn't as easy to document in those days either. You had to have your secretary to type up the documentation. You could punch cards and cards would serve as some kind of documentation. It was a different world. Eventually big modelers came around, and I credit them, even though their realization came late, by the early 90's they finally got the message that you can't just have a Wild West attitude. You really need to be careful writing scientific software, you have to document it properly, you have to configure and control it, and you can't just randomly make changes to it when you find a bug. Even when you find a bug, don't just go and fix it because if you find one, you will find another one. The philosophy was, oh gosh I have found the bug and if I fix it, everything will be wonderful. But the software engineers were right often there were more bugs, more subtle ones, so they go over this endless cycle of core dumps, looking for the error and fixing another bug. It did not occur to them that there was a more rational way to develop scientific programs.

**You wrote the NCAR tech memo in 1979 about MIEV0, MIEV1 and revised it in 1996 and these programs are still widely used. There are 700-800 references to one of your papers on Mie code alone. In that report you list several goals: maximum speed, generality, reliability, avoidance of numerical instability, portability, accuracy, and as simple and**

**straightforward as possible. This was in 1979. Which of these are you still finding appealing to you and why? To which extent "elegance" of the program was of importance in your work?**

Did I write it? It is beautiful. I think I know what you are driving at. It is certainly unique to me. I migrated to the field from applied mathematics and I had a mathematician's take on things. Mathematicians are big on elegance. The big accomplishment is to prove something in fewer steps. I felt that those same ideas can apply to computer programs and I did the best I could to make them simple and elegant. It was not something which was around at the time. If I said something like that - people would laugh. Computer programs were just a means to an end. Not something beautiful in themselves. I never took that attitude. I have always thought – make them as beautiful as you can. It was my background which made me trend towards elegance. Not often something you hear. Software engineers do talk about it, maybe not in such reverential tones as I do.

**Do you think that elegance helps programing? Makes it less cumbersome to debug the program?**

Yes definitely. Often I would make cosmetic changes to programs. At least in my view it would make the program more readable, more robust against people introducing errors into it. This is always the danger that I was aware of. People can take your lines of code and introduce errors and I was always aware of it. I was aware that maybe if you make it simple enough maybe they will not introduce errors when they start fiddling with it. To me this was something deeper.

**I have asked you also about maximum speed, generality, reliability, avoidance of numerical instability, portability, accuracy. I have a reason for it. We are often driven by hardware, in that time it was vector processing.**

Maximum speed is not on my list anymore. I guess that reliability and robustness are the most important. I was very much influenced by the guy in software engineering who said that one has to try to break the software. He was very much of the philosophy that we are too soft on software. We should be tough on software and make it break, because if do not try it we will never know how robust it is. For me reliability is very important. Elegance would be secondary priority. If the code is reliable, well documented, and of course not having numerical instabilities, but that is almost a given.

My interest was to take given equations and to do the very best job with them. I wanted to produce the very best result for those equations. When I worked with Nussenzveig part of the work was to undermine Mie calculations because he was saying that for larger particles one can use other formulas. But even there I was stickler for detail. Getting Airy functions just right, getting Bessel functions just right. I would write my own routines. I would not necessarily rely on numerical libraries. I wanted all special functions tested and reliable and failure tested. I pushed them as far as I could and they did not break. Or maybe they would break and then I would fix them.

**What was your involvement in research with Nussenzveig and software for spheres with very large size parameter? What was the motivation for this research?**

I was frustrated. Mie scattering calculations for large size particles were just horrendously expensive and most people just defaulted to geometric optics which would kick in at size parameter of 1000, although you can argue 500. Really there was a no man's land of size parameter between 100 and 1000 where the calculations were just bloody expensive if you were doing integration over sizes so I was always on lookout for ways to do that better. Just at that time, I became aware of a student of Nussenzveig named Khare. So I went to Bob Dickinson and said, let's invite this guy Khare. He said sure, invite him. When he arrived to NCAR he said – oh, by the way, my thesis advisor Nussenzveig is in the country visiting from Brazil. So I went to Bob Dickinson and he said – invite him too. We paid for their travel and the rest is history. In the 1970s at NCAR you had an almost unlimited travel budget; you only had to justify it to your branch chief. This was immensely productive and kept the atmospheric community tightly coupled.

Nussenzveig and I hit it off and I became his super programmer. I brought everything I knew about scientific programs. He needed that because he did not program at all. He was an ivory tower guy. He derived these long formulas, but he did not have a clue how to calculate them. He and I were just perfect for each other. I knew how to calculate stuff and he knew how to derive stuff. I was not entirely uninvolved with it, I would argue with him. Can we improve this, can we improve that, we had some interactions on derivations but he was definitely leader on that. I was entirely responsible for writing programs and I was very careful. We produced a string of papers. The last one we did was on resonances which bring us full circle because he actually developed formulas which were predicting where resonances were. You could avoid them if you want to. I am not sure if the calculations of where they are would be so expensive that it would not be worth it, but in principle you could avoid them.

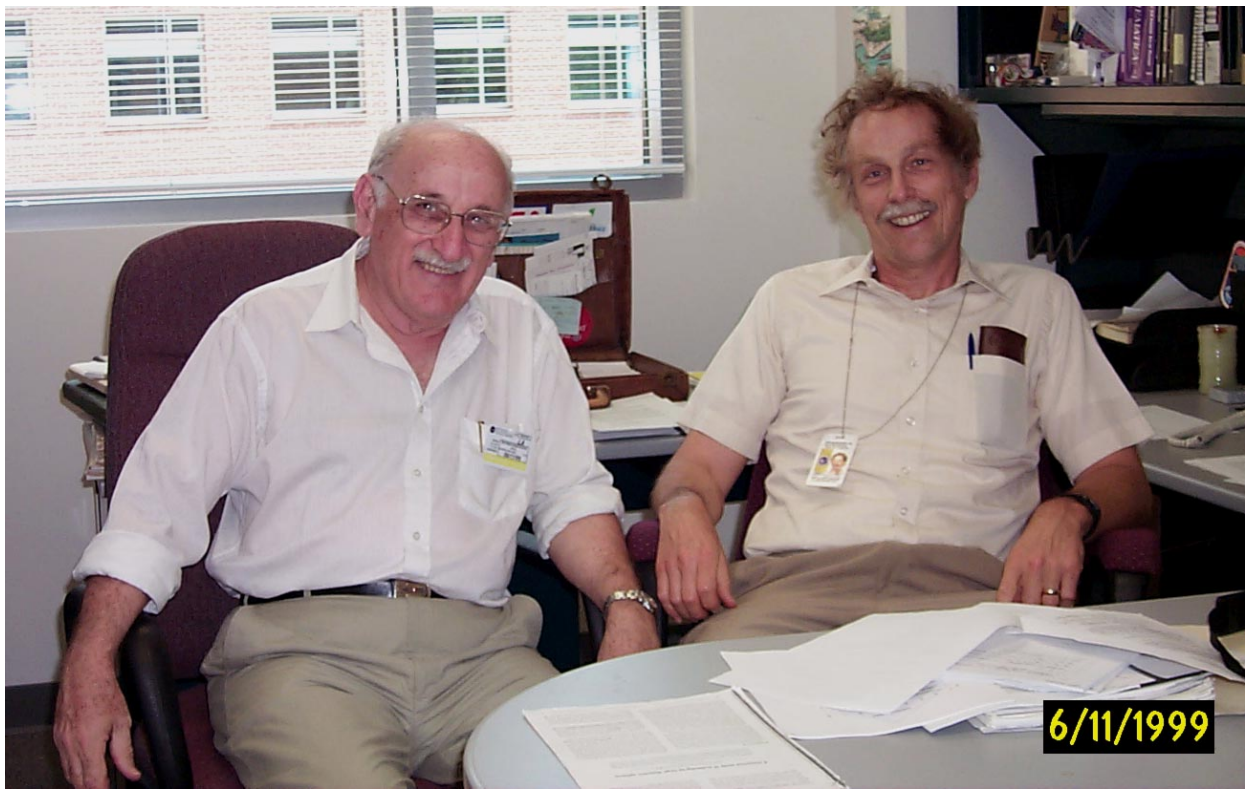

Wiscombe and Nussenzveig, NASA, 1999

**One of the theories which were used was called CAM – Complex Angular Momentum. Could you comment on this approximation?**

CAM theory was an offspring of a theory of 1950' which was called Regge pole theory [Regge]. Nussenzveig took a look at Regge pole theory which was quantum mechanics theory and he just grabbed it and dragged it to classical physics and applied it to Mie scattering problems. He also brought some work from Debye's thesis [Debye]. Debye said that there is an alternative to Mie expansion; you can expand in a different way. Nussenzveig learned to calculate each term of the Debye expansion in his new Complex Angular Momentum theory and it was novel, everybody agrees. This made him such a great physicist. It was really cool. He married his own work with the work of Debye which just sat there. Van de Hulst mentioned it in his book, of course. But it kind of sat there, and he really expanded this work and made it really relevant. Now, we were actually able to calculate terms in the Debye expansion, which converge very quickly because it amounted to expansion in terms of multiple internal reflections and there were these pesky surface waves terms which we had to add in. Those were where most of the trouble was; above edge and below edge rays - they kind of skitter along the edge of the sphere. It was a very physical way of looking at scattering. Mie series themselves doesn't give you much physical insight except in Rayleigh limit. Debye expansion gives you quite a little bit of insight, although when you get to surface waves you still feel like you skate on thin ice. But they are there, you can measure them. It is a beautiful phenomenon.

**Was the CAM theory practically applicable?**

Definitely. I could not have done the large droplets paper without CAM theory.

**Oh, really.**

Even then in the early 80's Mie calculations were so expensive. I remember one night I used eight hours of CRAY time and I was called on the carpet by Chuck Leith who was the head of NCAR science at the time and he asked "What were you doing burning eight hours of CRAY time on Mie scattering calculations" and I said but … it just takes that much time because particles are large and series is long, there is a lot of different sizes you have to integrate over. I don't think he was convinced. Anyway, it was hard to do large drops. By which I mean drops bigger than 10 microns (the favorite size of radiation people) but smaller than drizzle droplets. Sort of a no man's land of drop sizes. Those were very hard to calculate.

**Did you use it when you were working on "The effects of very large drops on absorption" paper? This is the first entry I have in my folder of your papers. It is in front of me.**

Yes. My earliest papers with Nussenzveig came out in the early 1980's and after that we collaborated on several other ones thanks to NASA Postdoctoral Fellowships for Nussenzveig at two different times. We even did a paper on bubbles, which you were interested in. Now they are coming back – people are studying bubbles in the ocean for geoengineering.

> **Efficiencies for Radiative Transfer in Atmosphere**
>
> Efficiency $Q \equiv (\text{cross section } \sigma)/\pi a^2$
>
> $$Q_{ext} = \frac{4}{\beta^2} \operatorname{Re} S(\beta, \theta=0) = \frac{2}{\beta^2} \sum_{\ell=1}^{\infty} (2\ell+1) \operatorname{Re}(a_\ell + b_\ell)$$
> ↑ optical thm.
>
> $$Q_{abs} = \frac{1}{2\beta^2} \sum_{\ell=1}^{\infty} (2\ell+1)\left[(1-|S_\ell^E|^2)+(1-|S_\ell^M|^2)\right]$$
>
> $Q_{extinction} = Q_{absorption} + Q_{scattering}$ ($\leftrightarrow \sigma_{total}$)
>
> $Q_{pr} = Q_{ext} - \overline{\cos\theta}\, Q_{sca} \equiv$ Radiation pressure efficiency
>
> $$\overline{\cos\theta}\, Q_{sca} = \frac{4}{\beta^2} \sum_{\ell=1}^{\infty} \frac{\ell(\ell+2)}{2\ell+1} \operatorname{Re}(a_\ell^* a_{\ell+1} + b_\ell^* b_{\ell+1})$$
> $$+ \frac{4}{\beta^2} \sum_{\ell=1}^{\infty} \frac{2\ell+1}{\ell(\ell+1)} \operatorname{Re}(a_\ell^* b_\ell)$$
> (P.J. Debye, 1909)
>
> $\beta \gg 1$ Results from geometrical optics + class. diffraction
>
> $$Q_{ext} = 1 + 1 - \frac{8N^2}{(N+1)(N^2-1)} \sin[2(N-1)\beta]$$
> ↑ geom. opt.   ↑ forward diffraction peak   ↑ transmitted axial ray   (N real)
>
> $$(Q_{abs})_{g.o.} = \int_0^{\pi/2} d\theta \sin\theta \cos\theta (1-e^{-A}) \left( \frac{1-|r_1|^2}{1-|r_1|^2 e^{-A}} + \frac{1-|r_2|^2}{1-|r_2|^2 e^{-A}} \right)$$
> (H.C. Van de Hulst, 1957)
>
> $\theta = \theta_1$ (∡ of incidence)
> $r_j = r_j(\theta) =$ Fresnel reflectivities ($j=1,2$)
> $A = 4\kappa\beta \cos\theta_2$, $\sin\theta_1 = n \sin\theta_2$, $N = n + i\kappa$
> Similarly for $(Q_{pr})_{g.o.}$   Results not accurate enough!

Nussenzveig handwritten derivations.

**<span style="color:red">Oh they are?</span>**

Yes, people are talking about modifying clouds by making huge bubble clouds in the ocean, which eventually produce cloud condensation nuclei when they come to the surface and pop.

**Interesting. I actually wrote a paper about bubble clouds in which I have classified them into bubble cumulonimbus, bubble stratocumulus and I calculated their radiative properties. Anyway, let's go back to you. You put together a paper on "Scattering by Chebyshev particles" with Alberto Mugnai who was in that time also in Fort Collins.**

Yes, and he was at NCAR for a while as a postdoc. He seemed bright to me. I was curious about the phenomena of non-spherical scattering and I realized that Mie theory was limited to spheres or near-spheres and I thought let us fly in the dark and try to adapt this Waterman's method which at that time was not called T-matrix. It was called EBCM – extended boundary condition method. How do I even remember it? Alberto was really good with computer code like me. He basically took some old EBCM code, may be got one from Waterman or may be not. He developed it like crazy, he made it really good. We used this code much, we just beat it to death, all the work we did after that. I was really interested in the effect of concavity, what happens when particles are concave, are there trapping modes? Is the radiation preferentially absorbed when there are these concavities? What is going on? Maxwell equations ought to be able to provide answer to that. It turned out to be devilishly difficult because if you make particles too concave the EBCM method doesn't converge. You realize that it is probably not converging series. It is an asymptotic series. It is series which when you sum it up to certain point it is good, but when you sum beyond that point it gets worse. We never could do anything too concave but we did manage to get mild concavity using this Chebyshev shape. The Chebyshev thing came from the fact that I was an applied mathematician and I have loved special functions. I looked at my Abramowitz and Stegun [AbramowitzStegun] and looked around for a function which would make sense to apply to non-spherical particles and I said let us do the Chebyshev thing. This is how it happened.

**Very recently I have used your Chebyshev ideas. I worked with a student at Scripps Institution of Oceanography on modeling of internals of a realistic biological cell. I remembered your Chebyshev particles and I told him to use it. It was like a year ago. I gave him your paper.**

Oh, that's amazing.

**How long it lasted, your love with non-spherical particles?**

It went on over 10 years. The reason was partly personal. Alberto and I got along quite well and I really enjoyed going to Italy. At the time I did not have anybody funding this research. Alberto managed to provide the money. He paid all my travel and NASA said go. I would go over there for like a month and we would do a paper together. He would make me teach something when I was there. One time when I went over I taught scientific software which relates to our conversation. I said to him that I am just bored with all that Mie scattering, I want to teach something different. That was the first time I ever did it, in Italy. It is funny how things develop; as a result of going there to him I developed my thoughts in writing, although I have never published a book. I am very sad about it. I think that I should have written a book about scientific software. I had an extensive set of notes which I distributed for years. It was called – writing scientific software. A lot of people seem to still know about it, at least as of 10 years ago people would still ask me about copies of it, and it all came out from this course with Alberto.

**Alberto Mugnai. He was in Rome. I think I visited that place once. I knew Gianni Dalu who was from there and he is my friend. It was not University of Rome; it was more like academy of science, right?**

*To Istvan, with best regards,*
*Warren*

CONSIGLIO NAZIONALE DELLE RICERCHE

# ISTITUTO DI FISICA DELL'ATMOSFERA

PRINCIPLES OF NUMERICAL MODELING WITH EXAMPLES
FROM ATMOSPHERIC RADIATION
PART I

Warren J. Wiscombe

IFA 87/31                              SEPTEMBER 1987

Yes, it was a research institute and I was teaching there. He just rounded up his colleagues and made them come to my lectures on scientific software. I told them that it was good for them, that they probably wrote crappy code.

**What is your view on the role of computers in light scattering applications? Without them we would not be able to calculate many interesting phenomena such as resonances. On the other hand we would probably have many theoretical developments such as the anomalous diffraction theory of van de Hulst or CAM? In atmospheric sciences we have similar dilemma. For example, one wonders if we would have semi-geostrophic approximation if we had powerful computers at that time. What is your take on this?**

I tend to agree that computers undermine thinking about simple approximations; that is certainly a loss. On the other hand you may say we had many years. One can start the clock from Maxwell equations, but say start the clock from Mie theory. We had more than 50 years to the time of van de Hulst's book. May be they pretty much exhausted what was possible. The best new approximation was Complex Angular Momentum. Those came along in the 50's and even those were rather complex. This was first breakthrough that came along in several years. Perhaps there weren't any more simple approximations. Maybe computers did not destroy the progress of simple approximations. The same thing has happened in quantum mechanics. You can't calculate analytically anything beyond some simple systems, say helium. I don't think that anybody in that field thinks that if they had withheld computers they would now calculate analytically lithium atoms. Computers were the next logical step. Another thing which I would like to add is something about which I disagreed with my thesis advisor Gerald Whitham of Caltech. He believed in closed form solutions. Say, solutions to partial differential equations. But those exact solutions would be triple sums over a bunch of Bessel functions and other things. I would say to him, look if I can compute this directly using difference methods, how is that worse than trying to sum up this triple sum of Bessel functions? We never agreed on that. But I think my attitude was more modern. Either way you have to do a boat load of computing to get an answer even if you have a so called analytic solution.

It is sad that scientists these days don't get as good an applied mathematics education as they need. I got a pretty good education in applied math, you did, and many people we worked with had a good applied math education. That seems to be dying now. I think that not having a good applied math education, not knowing your way around the modern version of Abramowitz and Stegun (the one authored by Frank Olver and others), is a loss. Depending too much on numerical solutions without knowing what is under the hood will lead to scientists who know how to pull the levers but who don't know how the machine works.

**I agree. I tend to tell my students that there is nothing magical even about sin(x). After all it is calculated using a series approximation. The simplicity of sin(x) is artificial. Sometimes triple series are more complicated to calculate than an integral representation which can be done by quadrature. It is interesting to hear you to say all these things. It is almost as if they come from my mouth.**

My final comment is that blending of computational and analytical points of view are the most powerful. One example EBCM; now everyone is calling it the T-matrix method. This is a nice blending of these two methods. It is not just a brute force solution. It is clever. That is where we should be heading. When I was developing my codes I made use of my applied mathematics knowledge, about series, convergence, asymptotic series, and all this stuff which I learned in grad school.

**Your light scattering publications have been referenced hundreds of times. What do you think is your lasting legacy in attacking light scattering computational problems? Are we just passing the torch and there will be better methods after us, or perhaps our legacy is not purely driven by hardware speed?**

It is a very deep question. As far as my legacy I will let other people to talk about. But I can talk about my philosophy. I did not just graduate with a Ph. D. with this philosophy, it took years to develop one. I was influenced by many people, notably by Freeman Dyson. I have the tool making philosophy. I believe that I created some pretty good tools that other people used to good effect. That makes me happy. I would regard these tools as part of my legacy. What Freeman Dyson argued, and I don't know anybody who refuted this, is that the role of tools in science is larger than the role of new ideas. He says that in any given point in time there are a lot of new ideas kicking around. There is never a shortage of ideas. But there is always a shortage of good tools. He said that if you look at the history of science it takes great leaps when new tools come along. Of course, we all know the story of the telescope and Galileo but he goes on to tell other stories such as the electron microscope, the original microscope that allowed Pasteur to discover the bacteria. His point was that we vastly underrate the importance of tools in the development of science and we overrate the value of ideas. I am very much in the tool development business. I like to develop tools. In fact I look at ARM as a tool. I helped to develop ARM and while I would never tell ARM folks that – I looked at it as a tool. We provided a tool to the community. They used it pretty well in my view. It gets back to the software tool philosophy. There are even books on these topics. Remember "Software Tools" [SoftwareTools] written by the guy who invented UNIX [Kernighan] Kernighan.

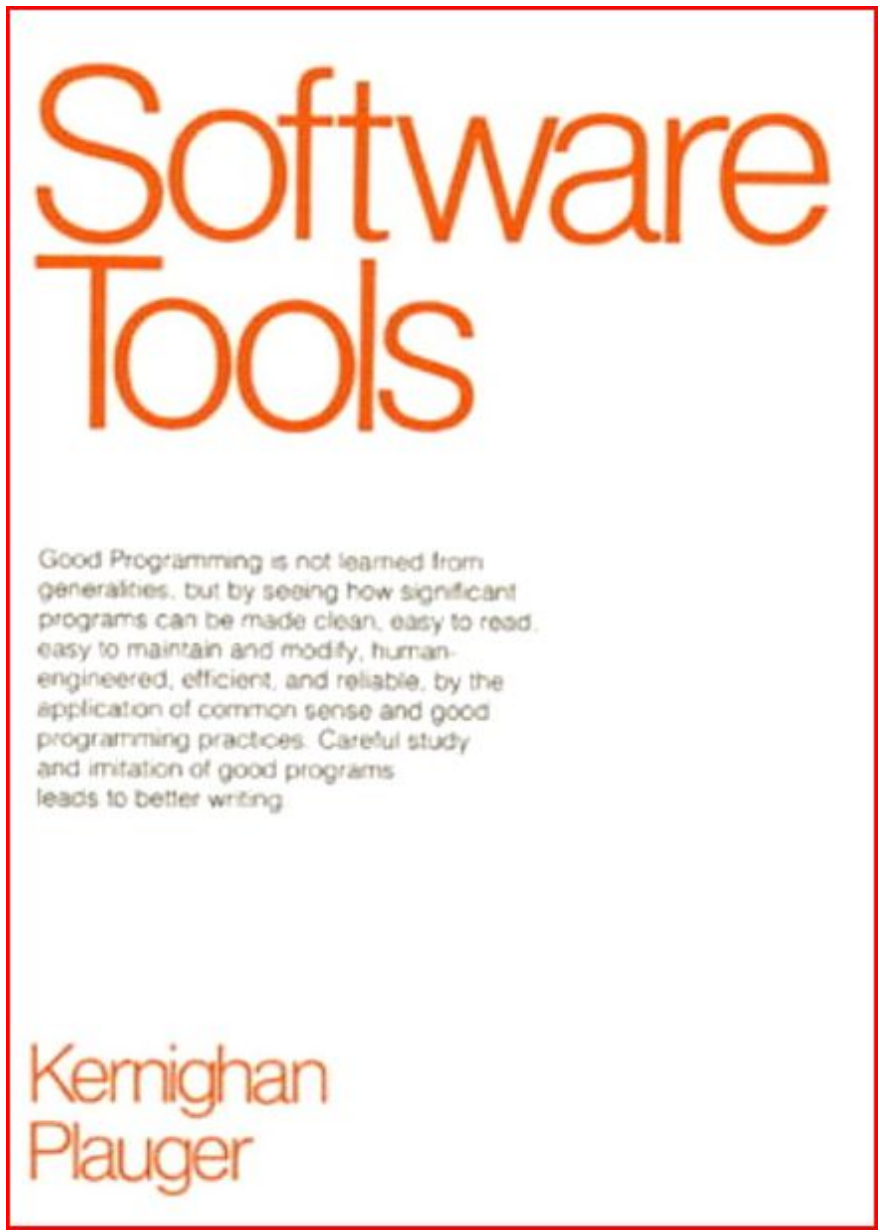

**I have read it.**

They wrote two books that influenced me. One was called Software Tools in Pascal, but the principles are very general, not just applicable to Pascal which no one seems to remember.

**I do. I remember them.**

The other one they wrote was The Elements of Programming Style which is a very thin book. I read it, I pored over it, scribbled on every line, and it is like a Bible. I learned many of the things which I subsequently applied from that book. It is a very nice book.

**You think that in light scattering your role was that of a toolmaker?**

Yes. I was a facilitator. I facilitated a large amount of other research. That pleases me. I like that. I would be happy to be called the toolmaker, among other things.

**We have already talked for an hour. Are you OK or are you tired?**

I am OK. By the way I have stolen some of the words from the Software Tools book. I just see that on the title page it mentions some of the words we discussed before. I was a disciple of those guys.

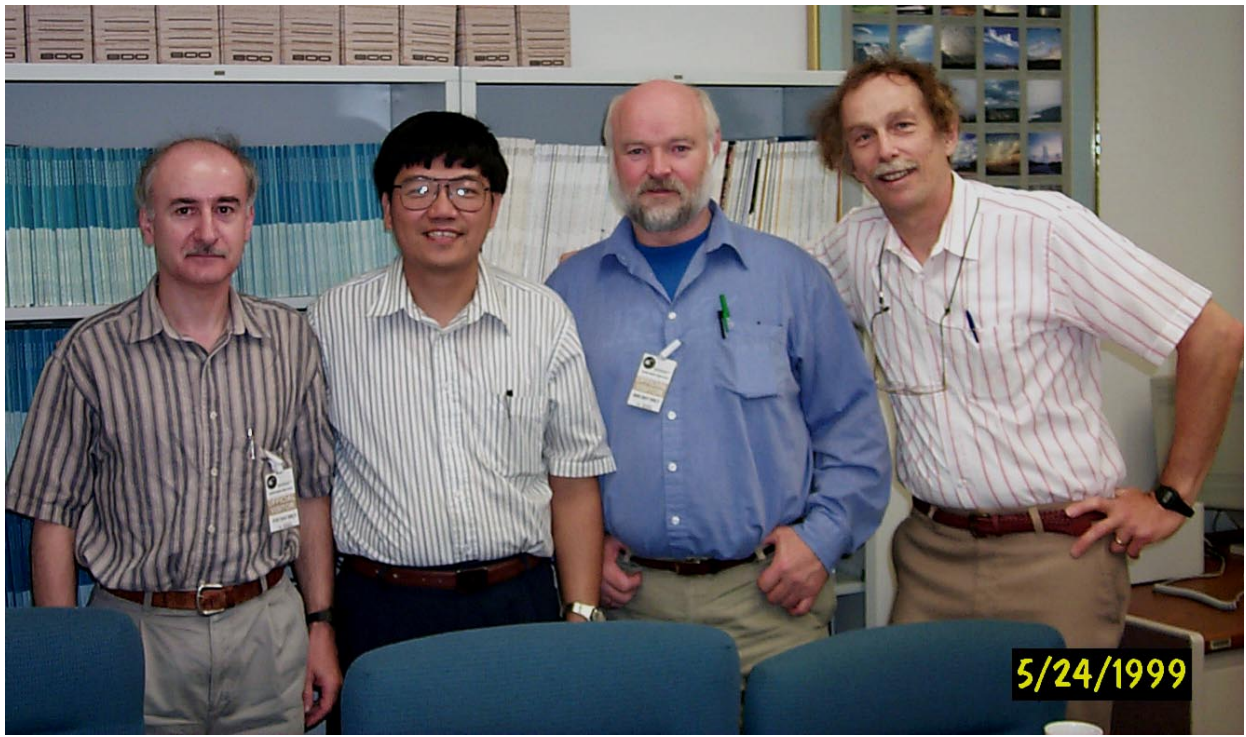

From left to right: Istvan Laszlo, Si-Chee Tsay, Knut Stamnes, warren Wiscombe. DISORT gang.

# Radiative transfer

[Piotr J. Flatau] **Let us move on to your other achievement, this time let us talk about your radiative transfer contributions. Your 1976 delta Eddington paper was followed by a series of reports and publications providing us with two stream radiative transfer solvers. I think these opened a way for the 1980 review by Meador and Weaver and even current two stream codes which benefited from your publications.**

[Warren Wiscombe] This was my first big hit. I do think that we sparked a lot of activity in that area. It was kind of a dead field. What I saw was climate coming along very strongly as a subject. This was in the days when dynamics dominated the field. Of course there were not that many of us. I was strong believer that eventually they have to recognize that radiation drives climate and that they will need a better radiation package. The radiation packages they had in their models were awful or nonexistent. They had radiation packages in which Sun never rose, they just used diurnal averages. They were very dismissive of the radiation. I knew that they will need better radiation and sure enough not long after delta Eddington came along there was really quite a move on the part of climate modelers to get better radiation packages in the shortwave but in the longwave too. When they were working with shortwave they really needed something like delta Eddington and I made it easy for them to implement it by publishing a report of how to do it numerically and a code as well. I do think that I stroked a cord there. This frantic activity up until 1980-1990 and the Meador-Weaver paper was just the realization of other people that it was needed. There was a lot of competition. You would not have been aware of that. Like the French. They thought they had a better approximation.

**The Lenoble group in Lille. Tell me more about 1976 paper in terms of numerics.**

Joseph came to visit NCAR. He had been in Wisconsin working with Jim Weinman and he was again kind of a Nussenzveig kind of character. He was from the older generation. It is humorous for me to say it now. He did not do computers. He just did not take to it the way we did. I was perfect for him. I said, you have an idea but an idea is worth nothing unless you test it. Let us test this thing the way no one ever tested it. Let us compare it with the exact calculation which I was able to do because of ATRAD. Let us test delta Eddington, see if we can make it break. We did huge amount of calculations, only a few appeared in the paper. We tried to break it and we determined what its limit of applications is. That was breakthrough. We did not just publish a few formulas and say, well we think it might be a better approximation but we really tested it thoroughly. This is why we got so much attention. Because people could really put some trust in it. It was adopted in many GCMs over the time. Now GCMs have a variety of things, whether it is Fu-Liou. There was a time when delta Eddington pretty much dominated GCMs. Tony Slingo adapted it for the British Met Office.

**In your report you used a pentadiagonal solver. Was that new or in that time people knew how to apply tridiagonal and pentadiagonal solvers to two stream?**

They were aware of how to do it. I was lucky that I started my work when numerical analysis was just exploding. LINPACK, NetLib, the whole idea of sharing codes was coming into being as contrasted to previous times when people did not share codes. I had an applied math background so I was comfortable working with matrix theory. It was like mother's milk to me. I was a perfect fit. I hammered on the math – formed it to shape and hopefully made it stable. But I still remember instabilities I couldn't control -- there were some IF statements in the delta Eddington code. Since I didn't know what to do, I just said "If this condition occurs, stop".

**This is a more scientific question, but I am curious. Were you able to do molecular properties or at that time you were mostly concentrating on the two stream solver itself?**

I was interested in actual fluxes because I was always climate oriented. For me fluxes were the number one item. I was not really into remote sensing for quite a while. I was a flux person and I hated the name change from "flux" to "irradiance" because it implied that radiation was somehow a different kind of energy than sensible or latent heat.

**You had both cloud and molecular properties integrated within the model, right?**

As far as molecular properties, you may remember, the only thing around was LOWTRAN. We all used LOWTRAN. It was not very good compared to what we have today but it was nice in that everyone used it (and cursed it at one time or another). In fact I rewrote LOWTRAN from scratch. I got the code in 1972 from Bob McClatchey at AFCRL. The code was so terrible that I went to my supervisor at that time, Burt Freeman, and I said – I can't use this, it is just awful, so I need to rewrite it. He said, go ahead. He gave me a couple of months and I rewrote LOWTRAN from scratch and added a lot of documentation in the form of comments. Actually I distributed it for a while because people became aware that I had a version of LOWTRAN which was robust and it had documentation and it was not full of GO TO statements and weird COMMON blocks. I was a secret, under the table, LOWTRAN distributor for a while.

**I did not know about this one.**

I should dig it out at some time. Even my version would look pretty pathetic by modern standards. The original version was what we called "spaghetti code" – so tightly wrapped around itself that you couldn't find your way out unless you left a trail of popcorn.

**In 1976 you published the Delta-Eddington report. How well was it received?**

NCAR was quite good at distributing it. In those days an NCAR report meant something. It got good distribution. It hit the market that was ready for it.

**Do you feel that current two-stream codes benefited from your work?**

There was friendly, sometimes not so friendly competition. You have to remember that Fu-Liou is a spectrally integrated code. I never attempted to do it. My philosophy was to provide a good tool that is good for one wavelength. Let other people figure out how to integrate over wavelength. That is a different problem. I did the same with the Mie code. Even though I had a very nice code which integrated over a size distribution – I never provided it. I almost did at one point and I backed out - because people need to exercise their own creativity. Others developed codes that integrated over the wavelength and mine never did. ATRAD did, but ATRAD never got any wide distribution.

**I was the only one who was using it?**

No. Laszlo used it and Steve Warren used it.

**By the way I have now got an answer to my question about the whole package. You were never interested in the whole code for the two stream.**

Partly because I did not want to get sidetracked, there was a big problem for radiation people in those days, namely that dynamicists tried to pigeonhole you as a "parameterization person". That meant, aside from the diminution in your status, that you had to integrate over wavelength since climate models did not care about just one wavelength. I never wanted to get pigeonholed that way. I did not want to become a servant to some big GCM model. People would wind up in that capacity and maybe it wasn't so bad for them, since it seemed to guarantee lifetime employment, but I would have died of boredom. So I avoided developing GCM parameterizations like the plague. I was content to let other people to do that.

**That brings us to ATRAD itself. I asked you many years ago why you moved away from ATRAD and I know the answer, but I would like people to understand it. In that time you worked on adding and doubling. There were papers by Grant and Hunt [HuntGrant] and I recall that Graeme Stephens, who was my advisor, was using their approach as well in his Ph. D. thesis. Tell me more about ATRAD and why you moved from ATRAD to DISORT.**

ATRAD was funded by the ARPA Climate Dynamics Program. That program was ably led by Colonel John Perry. He was the one who very generously funded the development of ATRAD in the early days. It started in 1971. He supported me for 3 years until I went to NCAR. He was tough. He was asking me why are you doing this, why are you doing that. But he was satisfied with my answers and so he let me go on. I worked also on the AIDJEX sea-ice program and the stuff I did on multiple reflection between sea ice and cloud was all done with ATRAD.

John Perry was the real hero in terms of supporting ATRAD development. He gave me time to do it, to learn the field, to meet the big name players in atmospheric science and paleoclimate, to be able to make ATRAD comprehensive and bulletproof. ATRAD was quite a comprehensive model. Of course I loved the Grant and Hunt version of adding-doubling because it guaranteed positive radiances. Once you go over to DISORT, complaints still come in to this day "why do I get negative radiances?" The Grant Hunt method was really beautiful mathematically. I remember working through all the equations by hand, the whole thing from beginning to end, and I still have those handwritten derivations. When I implemented Grant and Hunt I did it from my own notes. I loved that method. I still think that it is a great method. But Stamnes slowly destroyed the arguments against Chandrasekhar's discrete ordinate method, one by one. There were many reasons why that was a bad method. I remember Liou in the mid-seventies, who was trying to use the discrete ordinate method and he published a paper basically saying that it is crap. He couldn't do anything with it. It may sound good on paper but it just doesn't work. Stamnes relentlessly demolished the barriers and made the method useful and that was very impressive work, with his various students. It just happened somehow that he invited me to come to Alaska in 1986 and I spent month and a half there working with him on the code and I convinced him that the code was something worth spending time on. He had a certain code and Si-Chee worked on it as a graduate student and it looked awful. It was really a mess. I said to him that he shouldn't just throw it all away and say that the papers are the only important thing. This code should be cleaned up and made available because it represents a really high level of development. We rolled up our sleeves and developed that thing. Later Istvan Laszlo showed up from Hungary. He came to visit me. He said, I would like to work for you, what I can do. I said - you are now in DISORT service. I trained him as well. The four of us worked to write the code. I did most of documentation and cleaning it up and developing the extensive test case suite. The other three they really did not know how to clean up the code. They just let me do it. At that time, it was rather unique to provide programs which actually tested DISORT so that when you changed it, you could determine if you did anything bad. Those testing routines took me quite a while. I am proud of them. They were nice routines. I made the whole package available for public distribution. I convinced them to do it and the rest is history. It was a very worthwhile activity. It was not that adding and doubling was bad. In fact I have often thought that I will go back, clean it up and run it again. Ha, I never got around to that. On my desk I have a circular wooden coin called a Round Tuit; on it is written "someday I hope to get a Round Tuit".

**That brings another question. Do you think that having dominating code is good for the field? On one hand it is positive to have DISORT around. On the other hand it blocks**

**development of competition. ATRAD is an example. I was always upset about you that you let it die. Do you think that dominance of one code may stifle the field, make it less progressive?**

I can't say that I did it for that reason, but I am not unhappy about it. I do believe that people should not take codes as black boxes. What they should take are pieces. The way you assemble a jigsaw puzzle. It is OK to take the Mie scattering piece. Not have to write it yourself. It is OK to take the Airy function piece and not to write it yourself because, I can tell you, it will take you a month. It is nice that these kids these days can assemble pieces, but they should learn to assemble a jigsaw puzzle together. If they don't ever do that, I don't think they ever get the idea of what modeling is all about. I am not unhappy that ATRAD did not get public distribution. I think it is OK. I am content with that.

**Well I am not. In 1989 Frank Kerr wrote a letter hiring Si-Chee Tsay. I remember Si-Chee from Fort Collins, we even wrote a paper together. In 1988 there was the DISORT paper. How has this collaboration developed in the last 20 years?**

One thing we have learned that I would tell everyone is, the better you document it the fewer questions you get. If you put the code out, you will save yourself a humongous amount of time if you document it. What we wound up getting was actually very few questions. People could just pick it up and use it, like a knife or fork. It was intuitive: you know what to do with the documentation. We created a package that we are very proud we did not get very many questions even though we knew that thousands of people were picking it up, downloading it. May be some of them were actually using it. That was one of the earliest things that we learned. Nevertheless there was fallout in terms of a steady drumbeat of questions that would come up over the years. They were not all dumb questions like why do I get negative radiances. Sometimes they were very subtle and deep questions. At that point I was involved with ARM and did not have time to deal with it. I got Laszlo to do it. Laszlo is the unsung hero of DISORT because starting in 1992 when ARM was just totally taking off and absorbing 110% of my time Laszlo stepped in and he supported DISORT. He would answer these questions. He would research what was wrong with the code; he would do small fixes. He was just great. I can't say enough about him. He deserves more credit than he got. He came from Hungary, hung around in NASA for a while, he could not even get in for a while so I would go out and meet him outside the gates and he worked with me for a while. He worked with Rachel Pinker at the University of Maryland and eventually he got a civil servant job at NOAA where he is now. In fact on Friday I will go to have lunch with him.

**Four of you - were you meeting from time to time?**

Knut was in Alaska. I had one time in Alaska and I went one more time. We worked for a couple of weeks and that was it as far as face to face with Knut. Si-Chee is of course at Goddard so I could see him more and Istvan was local so I could see him more often. We were early users of the internet. We would send versions of the code to each other.

1. Title: NUMERICALLY STABLE ALGORITHM FOR DISCRETE-ORDINATE-METHOD RADIATIVE-TRANSFER IN MULTIPLE-SCATTERING AND EMITTING LAYERED MEDIA
   Author(s): STAMNES, K; TSAY, SC; WISCOMBE, W; et al.
   Source: APPLIED OPTICS Volume: 27 Issue: 12 Pages: 2502-2509 Published: JUN 15 1988
   Times Cited: 1,598 (from All Databases)
   UC-eLinks

2. Title: OPTICAL-CONSTANTS OF WATER IN 200-NM TO 200-MUM WAVELENGTH REGION
   Author(s): HALE, GM; QUERRY, MR
   Source: APPLIED OPTICS Volume: 12 Issue: 3 Pages: 555-563 DOI: 10.1364/AO.12.000555 Published: 1973
   Times Cited: 1,585 (from All Databases)
   UC-eLinks

3. Title: PHASE RETRIEVAL ALGORITHMS - A COMPARISON
   Author(s): FIENUP, JR
   Source: APPLIED OPTICS Volume: 21 Issue: 15 Pages: 2758-2769 Published: 1982
   Times Cited: 1,500 (from All Databases)
   UC-eLinks

4. Title: LASER BEAMS AND RESONATORS
   Author(s): KOGELNIK, H; LI, T
   Source: APPLIED OPTICS Volume: 5 Issue: 10 Pages: 1550-& DOI: 10.1364/AO.5.001550 Published: 1966
   Times Cited: 1,484 (from All Databases)
   UC-eLinks

5. Title: TIME RESOLVED REFLECTANCE AND TRANSMITTANCE FOR THE NONINVASIVE MEASUREMENT OF TISSUE OPTICAL-PROPERTIES
   Author(s): PATTERSON, MS; CHANCE, B; WILSON, BC
   Source: APPLIED OPTICS Volume: 28 Issue: 12 Pages: 2331-2336 Published: JUN 15 1989
   Times Cited: 1,214 (from All Databases)
   UC-eLinks

**In January of 2012 there was 50th anniversary of Applied Optics. We met this year at the AGU conference in San Francisco and you mentioned that DISORT is now the most referenced paper published in Applied Optics. The code I developed DDSCAT is one of the**

**most referenced papers in Journal of the Optical Society of America. In fact I would like to beat you and have more references in the future. It is funny that two atmospheric scientists have these very successful papers in the premier optics journals in the world. Is this a source of satisfaction for you?**

Yes. That was one of the high points in my life when I heard that we are the most cited article in 50 years. I never in my wildest dreams would have imagined that. That is beyond cool.

**This is satisfaction of a toolmaker?**

Yes, I am very content about that.

## General issues

[Piotr J. Flatau] **My first question is about public domain. Both of us were successful in some sense. Both of us were interested in public domain releases, not many people were in that time. We got our prize but it took 20-30 years and it took a lot of effort. We had several codes which were successful and some of them were not. For example we wrote together a thermodynamics package [THERMOS] that nobody ever used or referenced. What prompted you to write public domain codes?**

[Warren Wiscombe] It was partly that Dave had made his code publically available. That was very impressive especially from the private company like IBM. That was kind of an example for all of us. I have always thought that that was wonderful thing. The second thing was development of the open source movement. That started back in early 1970' with invention of UNIX and C. I was swept out in that. In fact I became known as a distributor of open source software and I was once invited to the meeting in California on open source software and Richard Stallman [Stallman] was there. He spoke and I spoke about my experiences of sharing software. The movement to share software was in the air but it was not universally accepted. A lot of people thought that I was nuts to share my codes. They said you should keep them so you can publish papers. I said no, I think the best science happens when we share our stuff. I was very much in agreement with the open source movement that was going on. It was not clear that it was going to lead to a great career. A lot of people said that you will never have any great career by just publishing codes; you will never have any respect for that. Fortunately, I did not listen to them.

**I agree. I remember those times even now I feel that there are similar divisions on occasion. In 1996 you revised the MIEV programs and the NCAR note starts with "I didn't really want to do this. I have never liked backtracking and revisiting the same terrain twice." Is this what prompted you to attack so many diverse problems? Is it curiosity or getting bored?**

Yes, I would get bored. I have never liked to be pinned down. One of my favorite stories was about Richter, the guy who invented the Richter scale. Somebody mentioned to him the Richter scale once and he went ballistic. The lesson was that you don't want to be pigeonholed. You don't want to be known as Mie scattering guy or Richter scale guy [Richter]. It was always my thing, I wanted to be a variety of things, have a variety of experiences and try a lot of subjects. I wandered around some. It was not typical. Most people settled into a groove and they stayed in that groove for thirty years. You have seen it and of course I have seen it. I was always related to radiation somewhat and clouds, if it was not radiation.

**Your ftp site ftp://climate1.gsfc.nasa.gov/wiscombe/ contains many of your codes. The site says - this is a work in progress and is not, nor ever will be, complete ... I recall that similarly Donald Knuth, who wrote TeX, would have releases 3.1, 3.14, 3.141, 3.1415 … because he believed that his software will be perfect one day. Do you feel that some things are done?**

They are done as far as I am concerned. Other people are welcomed to develop them further. I kind of moved on mentally. I had an epiphany in the late 80's and early 90's that the radiation field was just dying in computer codes even though I was responsible for part of it. I used to say that we need to get these radiation people out from behind their computers and out into the field. I became notorious for saying this. I became this big advocate for experimental programs and ARM being the most visible among the ones I had my hand in. I kind of moved away from writing software even though I was teaching this software course in mid 90's. These were retrospective courses, sort of I lived this life, that is what I have learned. You can do better scientific software than you think you can and here is some ways you might be able to do that.

**I would like to check on you one of my theories. Often, when I review papers, and I was an associate editor of JAS for 7 years, I wanted to see the code. I wanted authors to provide the well documented code so I can check what they say. It was impossible of course. But what is your take on repeatability of scientific results these days?**

We used to discuss the repeatability of science. I think we lost it. We can whine about it but basically a lot of work which is published now with computer codes is not repeatable. You cannot really ask people to share their code. They may have 2-3 more papers to be published with this code and they don't want somebody else to take it. It would be difficult to expect reviewers to run the code. We lost the repeatability battle already. We can't get it back. I think it is sad. The way we try to replace it is with intercomparisons. We have these model intercomparisons, even measurement intercomparisons when we bring different instruments. We bring different models and run them with the same initial conditions and see if they give the same results. They don't usually, and we analyze the hell out of the differences. I think this is how we deal now with repeatability issue. Everyone knows it is a loss. Computer codes changed the game. They made repeatability and the old way of thinking obsolete. It is just gone.

**It is something that I did not think about, something new that I have learned from you just now. We both are old timers and we mostly write in FORTRAN, I think. Both of us migrated through various FORTRAN releases Fortran66, Fortran77, Fortran90. You told me recently that you went to a Python class which I did also several months ago. Some people like to move to new languages, some don't. For example Bruce Draine, my collaborator of DDSCAT code, is very conservative and doesn't like change. What is your feeling about computer languages?**

I feel that we should be open to new ideas. I have always studied other languages. I have found languages which I liked more, for example Mathematica, and languages which I liked less, for example C. C was too close to the bare metal of the computer, and too dangerous, and I could never see the need for C++ and its objects, at least not for hard-core numerical analysis and solution of physics and chemistry equations. Python looks interesting. The guy giving the Python tutorial said that it is an analysis language like Matlab and IDL, but free, and can't really replace FORTRAN. FORTRAN almost died of course. Until FORTRAN90 came along FORTRAN was really on a death spiral. It would not have lasted. FORTRAN90 breathed new life into the language. I think it will go on for many years now. It is very readable. It doesn't have weird symbols in it. When you use array syntax, the code kind of looks like the equations you write, especially when they are matrix equations.

**Did you switch your codes to FORTRAN90?**

I haven't been consistent on it. I played around but I was never consistent about it, shame on me. People volunteered to do it. I said – be my guest, but they have never finished their job. I haven't seen anything which is super superior to FORTRAN. Nothing that would make me want to switch. But I do think that other languages have their uses. Mathematica is fun. You can play around, plot Julia sets and plot all kinds of cool stuff which you would not imagine doing in FORTRAN.

**In 1979 you listed the FORTRAN code in the back of the Technical Note. It was common to do it. In 1996, in your revised version of the NCAR technical report the code was gone. We don't do this anymore. It is easy now to exchange code. What has changed in code writing in your scientific career which strikes you the most? Ease of software distributions, attitudes of the younger generation to distribute public domain codes?**

I would remind people that we used to punch cards. I have shipped to someone 2000 cards in a box and I have shipped 4000 cards. That was the way we exchanged codes. God help you if the card reader on the other end would not read your cards. The invention of the internet and ftp sites was just a revolution in terms of not shipping boxes of cards across the country. When I first started we used to copy the code. We would get listings of code and there was no way to get an electronic version – we would just copy it.

**I remember. You were running ARM which was very strong on data exchanged and codes which are exchangeable.**

It is like night and day. Attitudes changed; now people want to share codes. Attitudes have flipped. When we were younger people did not want to share the code. You and I were the exceptions. Now it is the rule. But it required changing the attitudes. It is easy to forget that. Everybody knows the technology. In that time LINPACK was a phenomenal success.

**When I was younger I would spend a lot of time on optimization of the code. Suddenly CRAY disappeared and with it vector processing. What appeared next was parallel processing and I stopped worrying about optimization because I wanted to concentrate on science, not on programing for hardware. Do you have a similar experience?**

Yes. I think that parallel revolution changed everything. Scientist tooks quite readily to vectorization. We have never adapted to parallelization. There were some horror stories in the early days. Like they rewrote the Mintz-Arakawa model for a computer called Illiac 4 and it was a disaster. It never worked. A huge amount of hours to rewrite it in some obscure language was wasted. Parallel machines were a watershed for scientists. Some crossed over; some did not. I did not. I dropped out partly because I was not able to cross to parallel machines. It required too much specialized coding. I believed in codes which are widely shareable which do not rely on specialized coding.

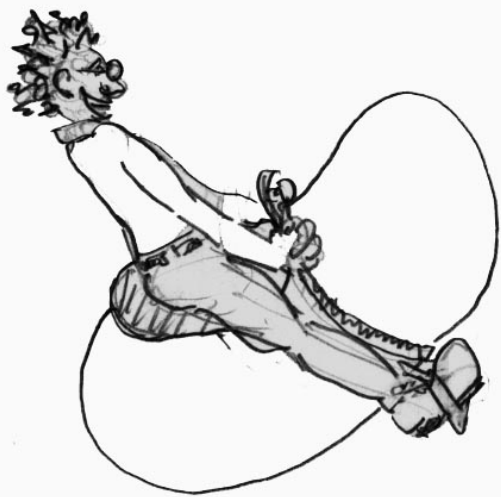

Wiscombe keep riding these big drops.  Artwork by Graeme Stephens.

**You have a PowerPoint presentation about "scientific revolutions" and mention in it that "sometimes the unresponsiveness of my colleagues to new ideas left me quite depressed". I know that you were discussing physical parameterizations. Do you think that codes can be an object of art, go beyond being tools and be revolutionary?**

Yes. They are, when they are elegant they are absolutely beautiful. Your last question is wandering into the realm of philosophy.  Maybe I was depressed that my colleagues are not sensitive to my new ideas -- in particular the large drop idea which I now like to crow about because they are finding drizzle drops in all kinds of clouds. At the time when I said that there might be a lot of drizzle drops in every cloud, cloud physicists were outraged. I still have a letter from a cloud physicist who dressed me down for suggesting drizzle could exist as a steady state condition for many hours without destroying the cloud. Who was little me to say that there are drizzle drops in all clouds or in many clouds?  Now ARM radars and Cloudsat are showing drizzle drops in vast numbers.

**The reason I pointed out this quote of yours is not to say that you are depressed. The reason was really related to programs. Do you feel that they can be revolutionary? For Mathematica was revolutionary in symbolic computing because all programs before just did not work.**

Yes, FFT is probably the best example. It is now in computer chips. It does stuff we even don't know about. It is beautiful code and look what it is doing.

**What about DISORT? Did it push the field beyond?**

Some codes show what perfection could be like in those times. Certainly it would not look as perfect now. Standards do improve. But it gave an example of perfection. Everyone needs to see these kinds of examples. I was reading you from the Software Tools in Pascal. I did not read you the last line. "Careful study and imitation of good programs leads to better writing." This is our long term gift. We show people, what at the time were perfect programs and they learn to write better programs as a result.

**Warren. It is 5pm on your time and we have talked for 2 hours now. I think we should finish. Thanks for all the answers. I agree with many of the things you said. There were some things you said I was surprised by.**

Thanks. Let me make my final points. The most important one is that you can actually write zero defect software. At that time for a scientist it was a revolutionary idea because we have always assumed that our software was buggy. There is no reason we can't write zero defect software. It is an idea from software engineering that we could learn a lot from. Another idea is that good software can live almost forever. In science our attitude always was that it is a throwaway. I have seen in my 30-40 years watching the field is that it is really true. Really good pieces of software have almost infinite life time. They just keep getting used.

**OK we are done. But just to make you happy. I am writing a paper with Jerry Schmidt about altocumulus and we observe drizzle size droplets there. We are reading your 1984 paper now about large drops and plan to reference it.**

Wow. Drizzle in altocumulus. Even I would be shocked.

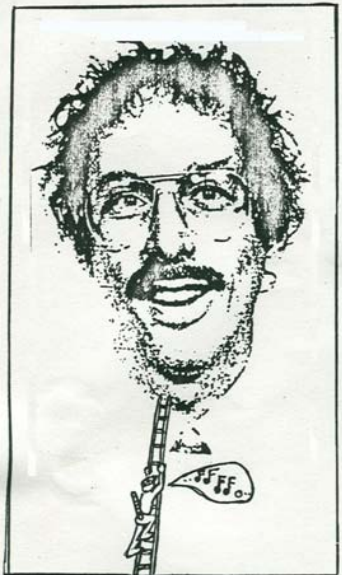
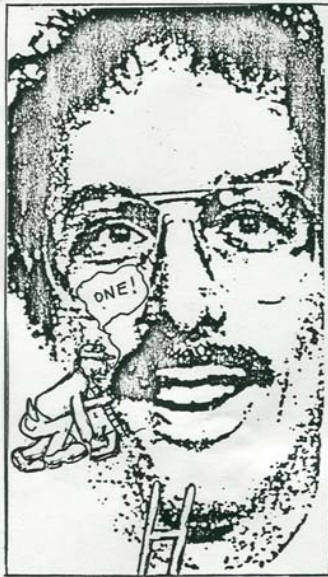
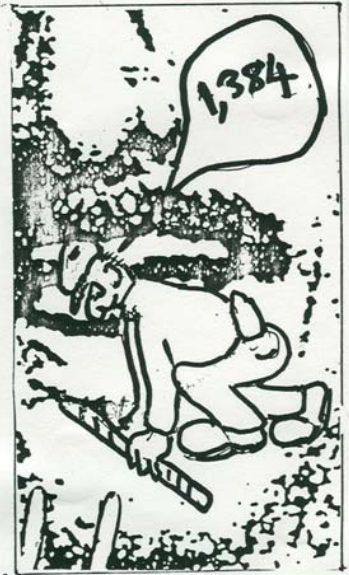
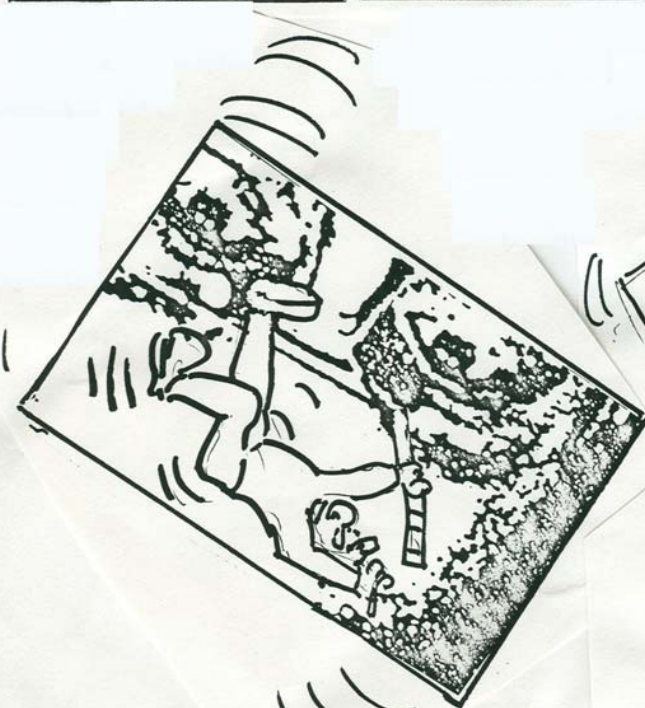
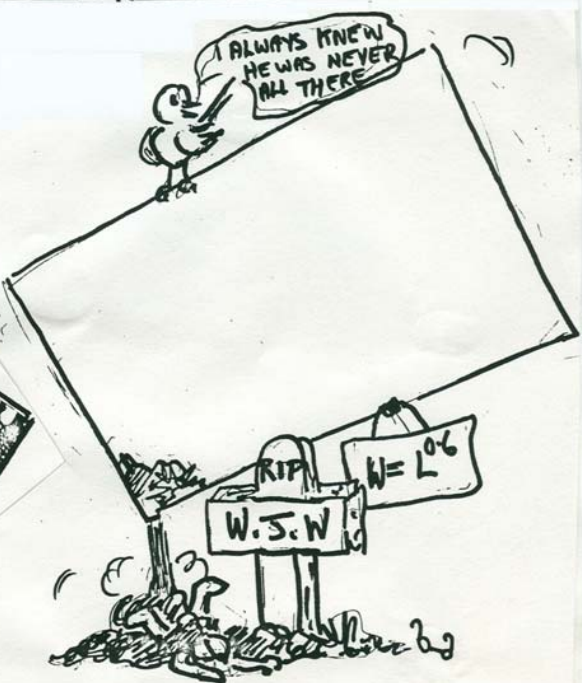

Wiscombe as fractal. Artwork by Graeme Stephens.

## Acknowledgements

I would like to thank Steven Warren, Istvan Laszlo, Aleksander Marshak, Graeme Stephens for providing additional graphics material and some suggestions.